\documentclass[twocolumn,trackchanges]{aastex63}

\newcommand{\HI}{\mathrm{H\,I}}
\newcommand{\lya}{Ly$\alpha$ }

\def\Mgii{Mg\,{\sc ii}}
\def\Alii{Al\,{\sc ii}}
\def\Cii{C\,{\sc ii}}
\def\Civ{C\,{\sc iv}}
\def\Siii{Si\,{\sc ii}}

\def\Feii{Fe\,{\sc ii}}
\def\Feiii{Fe\,{\sc iii}}

\def\Hi{H\,{\sc i}}

\def\Oi{O\,{\sc i}}

\turnoffeditone

\shorttitle{J0252--0503 Resides in a Significant Neutral IGM}
\shortauthors{Wang et al.}

\begin{document}

\title{\bf \large A Significantly Neutral Intergalactic Medium Around the Luminous \emph{z} = 7 Quasar J0252--0503}

\correspondingauthor{Feige Wang}
\email{feigewang@email.arizona.edu}

\author[0000-0002-7633-431X]{Feige Wang}
\altaffiliation{NHFP Hubble Fellow}
\affil{Steward Observatory, University of Arizona, 933 North Cherry Avenue, Tucson, AZ 85721, USA}
\affil{Department of Physics, University of California, Santa Barbara, CA 93106-9530, USA}

\author[0000-0003-0821-3644]{Frederick B. Davies}
\affil{Department of Physics, University of California, Santa Barbara, CA 93106-9530, USA}
\affil{Lawrence Berkeley National Laboratory, CA 94720-8139, USA}

\author[0000-0001-5287-4242]{Jinyi Yang}
\affil{Steward Observatory, University of Arizona, 933 North Cherry Avenue, Tucson, AZ 85721, USA}

\author[0000-0002-7054-4332]{Joseph F. Hennawi}
\affil{Department of Physics, University of California, Santa Barbara, CA 93106-9530, USA}
\affil{Max Planck Institut f\"ur Astronomie, K\"onigstuhl 17, D-69117, Heidelberg, Germany}

\author[0000-0003-3310-0131]{Xiaohui Fan}
\affil{Steward Observatory, University of Arizona, 933 North Cherry Avenue, Tucson, AZ 85721, USA}

\author[0000-0002-3026-0562]{Aaron J. Barth}
\affiliation{Department of Physics and Astronomy, 4129 Frederick Reines Hall, University of California, Irvine, CA, 92697-4575, USA}

\author[0000-0003-4176-6486]{Linhua Jiang}
\affil{Kavli Institute for Astronomy and Astrophysics, Peking University, Beijing 100871, China}

\author[0000-0002-7350-6913]{Xue-Bing Wu}
\affil{Kavli Institute for Astronomy and Astrophysics, Peking University, Beijing 100871, China}
\affil{Department of Astronomy, School of Physics, Peking University, Beijing 100871, China}

\author[0000-0003-2371-4121]{Dale M. Mudd}
\affiliation{Department of Physics and Astronomy, 4129 Frederick Reines Hall, University of California, Irvine, CA, 92697-4575, USA}

\author[0000-0002-2931-7824]{Eduardo Ba\~nados}
\affil{The Observatories of the Carnegie Institution for Science, 813 Santa Barbara Street, Pasadena, California 91101, USA}

\author[0000-0002-1620-0897]{Fuyan Bian}
\affil{European Southern Observatory, Alonso de Córdova 3107, Casilla 19001, Vitacura, Santiago 19, Chile}

\author[0000-0002-2662-8803]{Roberto Decarli}
\affil{INAF--Osservatorio di Astrofisica e Scienza dello Spazio, via Gobetti 93/3, I-40129, Bologna, Italy}

\author[0000-0003-2895-6218]{Anna-Christina Eilers}
\altaffiliation{NHFP Hubble Fellow}
\affil{MIT-Kavli Institute for Astrophysics and Space Research, 77 Massachusetts Avenue, Building 37, Room 664L, Cambridge, Massachusetts 02139, USA}

\author[0000-0002-6822-2254]{Emanuele Paolo Farina}
\affiliation{Max Planck Institut f\"ur Astrophysik, Karl--Schwarzschild--Stra{\ss}e 1, D-85748, Garching bei M\"unchen, Germany}

\author[0000-0001-9024-8322]{Bram Venemans}
\affil{Max Planck Institut f\"ur Astronomie, K\"onigstuhl 17, D-69117, Heidelberg, Germany}

\author[0000-0003-4793-7880]{Fabian Walter}
\affil{Max Planck Institut f\"ur Astronomie, K\"onigstuhl 17, D-69117, Heidelberg, Germany}

\author[0000-0002-5367-8021]{Minghao Yue}
\affil{Steward Observatory, University of Arizona, 933 North Cherry Avenue, Tucson, AZ 85721, USA}

\begin{abstract}
Luminous $z\ge7$ quasars provide direct probes of the evolution of supermassive black holes (SMBHs) and the intergalactic medium (IGM) during the epoch of reionization (EoR). The Ly$\alpha$ damping wing absorption imprinted by neutral hydrogen in the IGM can be detected in a single EoR quasar spectrum, allowing the measurement of the IGM neutral fraction towards that line of sight. However, damping wing features have only been detected in two $z>7$ quasars in previous studies. In this paper, we present new high quality optical and near-infrared spectroscopy of the $z=7.00$ quasar DES J025216.64--050331.8 obtained with Keck/NIRES and Gemini/GMOS. \edit1{By using the Mg\,{\sc ii} single-epoch virial method}, we find that it hosts a $\rm (1.39\pm0.16) \times10^{9} ~M_\odot$ SMBH accreting at an Eddington ratio of $\lambda_{\rm Edd}=0.7\pm0.1$, consistent with the values seen in other luminous $z\sim 7$ quasars. Furthermore, the Ly$\alpha$ region of the spectrum exhibits a strong damping wing absorption feature. The lack of associated metal absorption in the quasar spectrum indicates that this absorption is imprinted by a neutral IGM. Using a state-of-the-art model developed by Davies et al., we measure a volume-averaged neutral hydrogen fraction at $z=7$ of $\langle x_{\rm HI} \rangle = 0.70^{+0.20}_{-0.23} (^{+0.28}_{-0.48})$ within 68\% (95\%) confidence intervals when marginalizing over quasar lifetimes of $10^3\le t_{\rm Q}\le10^8$ yr. This is the highest IGM neutral fraction yet measured using reionization-era quasar spectra. 

\end{abstract}


\section{Introduction} \label{sec:intro}
The earliest luminous quasars, powered by billion solar mass supermassive black holes (SMBHs), can be used not only to constrain the physics of SMBH accretion and the assembly of the first generation of massive galaxies in the early Universe, but also to obtain critical information on the physical conditions of the intergalactic medium (IGM) during the epoch of reionization (EoR). Although more than 200 $z>6$ quasars have been found in the past few decades \citep[e.g.][]{Fan01,Willott10,Wu15,Jiang16,Banados16,Wang16,Matsuoka16,Reed17}, only several tens of them are at $z>6.5$ \citep[e.g.][]{Venemans15,Mazzucchelli17,Wang17,Wang19,Yang19,Reed19} and just six are currently known at $z>7$ \citep{Mortlock11,Banados18,Wang18,Matsuoka19a,Matsuoka19b,Yang19}. The limited number of known high redshift quasars is due to the combination of a rapid decline of quasar spatial density towards higher redshifts \citep[e.g.][]{Wang19}, the lack of deep wide-field near-infrared surveys, and the presence of  a large number of contaminants from Galactic cool dwarf populations in the photometric quasar selection process. Near-infrared spectroscopic observations of these known quasars indicate that billion or even ten billion solar mass SMBHs are already in place in these luminous quasars \citep[e.g.][]{Wu15,Shen19}. The existence of these SMBHs in such a young Universe challenges our understanding of the formation and the growth mechanisms of SMBHs \citep[e.g.][]{Volonteri06,Pezzulli16,Wise19,Davies19b}.   

Observations of the Lyman series forests in $z\gtrsim6$ quasars indicate that the IGM is already highly ionized by $z\sim6$ \citep[e.g.][]{Fan06,Bosman18,Eilers18,Eilers19,Yang20}, although the final completion of reionization might extend down to $z\sim5.5$ \citep[e.g.][]{Becker15,Davies18a,Kulkarni19,Keating20}. However, the Lyman series forests are very sensitive to neutral hydrogen and saturate even at low IGM neutral fraction (i.e. $\langle x_\HI\rangle \ga 10^{-4}$). On the other hand, if the neutral fraction is of order unity, one would expect to see appreciable absorption redward of the wavelength of the Ly$\alpha$ emission line, resulting in a damping wing profile \citep[e.g.][]{Miralda98} due to significant optical depth on the Lorentzian wing of the Ly$\alpha$ absorption. The first quasar with a damping wing detection is ULAS J1120+0641 \citep{Mortlock11} at $z=7.09$, although different analyses yielded different constraints on $\langle x_\HI\rangle$ \citep{Mortlock11,Bolton11,Bosman15,Greig17,Davies18b}, ranging from $\langle x_\HI\rangle\sim0$ to $\langle x_\HI\rangle\sim 0.5$ at $z\sim7.1$. Recently,  the spectrum of quasar ULAS J1342+0928 \citep{Banados18} at $z=7.54$ shows a robust detection of the damping wing signal \citep{Banados18,Davies18b,Greig19, Durovcikova19}, yielding $\langle x_\HI\rangle\sim0.2-0.6$ at $z\sim7.5$. Compared to other probes of reionization history, such as CMB polarization \citep{Planck18} and Ly$\alpha$ emission line visibility in high-redshift galaxies \citep[e.g.][]{Ouchi10,Mason18}, a main advantage of IGM damping wing measurement is that it can be applied to individual quasar sight lines, thereby constraining not only the average neutral fraction, but also its scatter in different regions of the IGM. However, the damping wing experiment is only feasible at very high redshifts where the IGM is relatively neutral, and current damping wing analyses have been limited to these two sight-lines due to the lack of bright quasars at $z\gtrsim7$. Thus, it is crucial to investigate the damping wing experiment along more $z>7$ quasar sight lines. 

In this paper, we present the detection of strong IGM damping wing absorption along the line of sight to a luminous $z=7$ quasar DES J025216.64--050331.8 \citep[hereinafter J0252--0503;][]{Yang19}, using new high quality optical/near-infrared spectroscopic observations; we also use the new spectrum to measure the mass and Eddington ratio of the central SMBH. In Section \ref{sec_obs}, we describe our photometric and spectroscopic observations for J0252--0503. In Section \ref{sec_bh}, we present the luminosity, BH mass and Eddington ratio measurements of J0252--0503. In Section \ref{sec_dp}, we discuss the reconstructions of the unabsorbed spectrum of the quasar and our constraints on the neutral fraction in the IGM at $z=7$ by modeling IGM Ly$\alpha$ absorption. Finally, in Section \ref{sec_sum} we summarize our results and briefly discuss the implications for the cosmic reionization history and BH growth constraints with larger quasar samples at $z\gtrsim7$ in the future. Throughout this paper, we assume a flat $\Lambda$CDM cosmology with $h=0.685$ \citep{Betoule14}, $\Omega_b=0.047$, $\Omega_m=0.3$, $\Omega_\Lambda=0.7$, and $\sigma_8=0.8$. All photometry in this paper is in the AB system.

\begin{figure*}[tbh]
\centering
\includegraphics[width=1.0\textwidth]{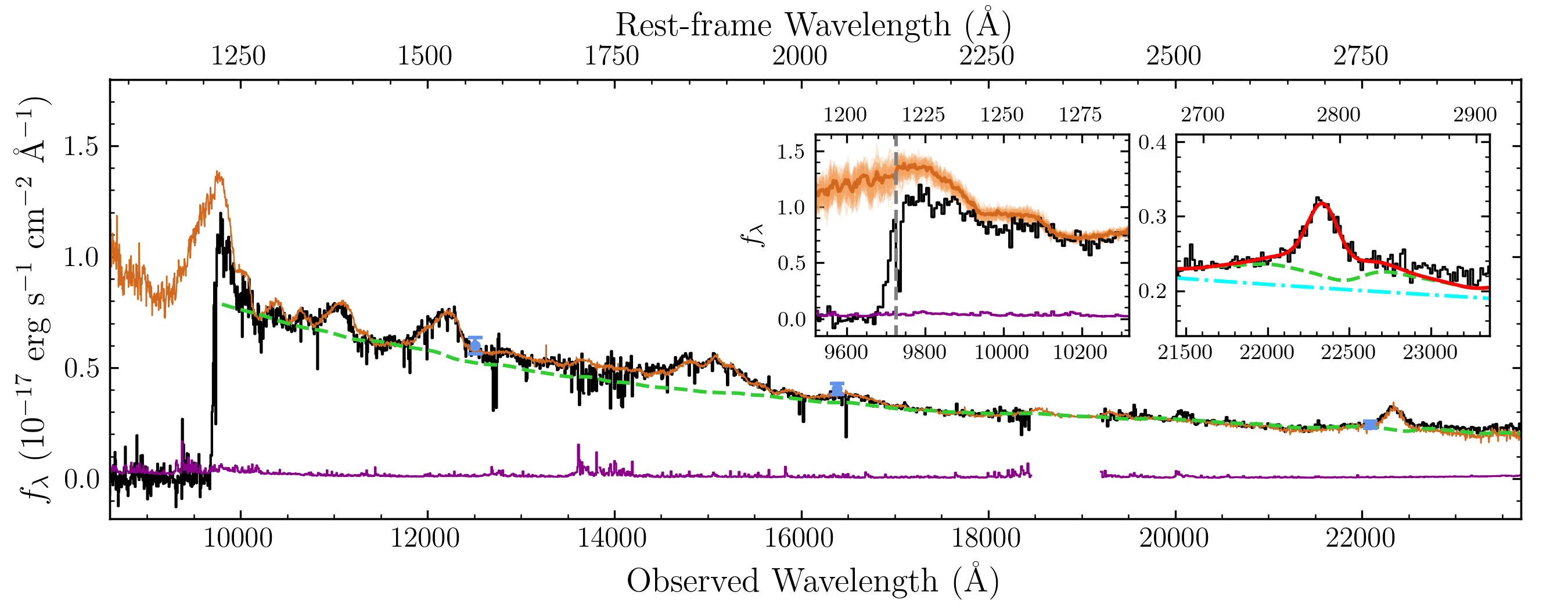}
\caption{Gemini/GMOS $+$ Keck/NIRES spectrum of J0252--0503. The spectrum is plotted using 200 $\rm km~ s^{-1}$ pixels (binned by $\sim 5$ native pixels). The black and magenta lines represent the Galactic extinction-corrected spectrum and the error array, respectively. The brown line denotes the quasar composite spectrum constructed with 83 SDSS quasars with similar \Civ\ blueshifts and line strengths. The green dashed line denotes the pseudo-continuum model which includes power-law,  iron emission, and Balmer continuum components. The light blue points are flux densities determined from Galactic extinction-corrected photometry in the \emph{J}, \emph{H}, and \emph{K} bands. The left inset is the zoom-in of the Ly$\alpha$ region. In addition to the composite spectrum derived from 83 SDSS quasars, we also show another 100 composite spectra constructed via bootstrapping. The Ly$\alpha$ position is marked with a gray dashed line. J0252--0503 shows strong absorption on top of and redward of the Ly$\alpha$ line, indicating a strong damping wing signature. The right inset shows the \Mgii\ line fitting with the cyan dot-dashed line denoting power-law continuum, the green dashed line denoting the pseudo-continuum model, and the red line representing total fit of pseudo-continuum and \Mgii\ line. 
\label{fig_spec}}
\end{figure*}

\section{Observations and Data Reduction}\label{sec_obs}
J0252--0503 \citep{Yang19} was selected as a quasar candidate using photometry from the Dark Energy Survey \citep[DES,][]{Abbott18} and the unblurred coadds of {\it WISE} \citep[unWISE,][]{Lang14} data. It was spectroscopically identified as a quasar at $z=7.02$ based on the strong \lya break using observations from Magellan/LDSS-3. However, the lack of a near-infrared spectrum for this quasar precluded detailed analyses in the discovery paper. 

We obtained a high quality near-infrared spectrum with the Near-Infrared Echellette Spectrometer \citep[NIRES\footnote{\url{https://www2.keck.hawaii.edu/inst/nires/}};][]{Wilson04} mounted on the Keck-2 telescope.  NIRES is a prism cross-dispersed near-infrared spectrograph with a fixed configuration that simultaneously covers the \emph{Y}, \emph{J}, \emph{H}, and \emph{K} bands in five orders from 0.94 to 2.45 $\mu$m with a small gap between 1.85 and 1.88 $\mu$m. The mean spectral resolving power of NIRES is $R\sim2700$ with a fixed $0\farcs55$ narrow slit. We observed J0252--0503 with NIRES for a total of 4.8 hours of on-source integration on three nights: 1.4 hours on 2018 August 12, 1.0 hour on 2018 September 3, and 2.4 hours on 2018 October 1 (UT). The observations were separated into multiple 300\,s or 360\,s individual exposures with the standard ABBA dither pattern. We also observed the flux standard star Feige 110. We reduced the NIRES data using a newly developed open-source Python-based spectroscopic data reduction pipeline \cite[{\tt PypeIt}\footnote{\url{https://github.com/pypeit/PypeIt}};][]{pypeit}. Basic image processing (i.e. flat fielding) followed standard techniques. Wavelength (in vacuum) solutions for individual frames were derived from the night sky emission lines. Sky subtractions were performed on the 2-D images by including both image differencing and a B-spline fitting procedure. We used the optimal spectrum extraction technique \citep{Horne86} to extract 1-D spectra. We flux calibrated the individual 1-D spectra with the sensitivity function derived from the standard star Feige 110. We then stacked the fluxed 1-D spectra from each night and fitted a telluric absorption model directly to the stacked quasar spectra using the telluric model grids produced from the Line-By-Line Radiative Transfer Model \citep[{\tt LBLRTM} \footnote{\url{http://rtweb.aer.com/lblrtm.html}};][]{Clough05}. Finally, we combined all spectra obtained on different nights to produce the final processed 1-D spectrum.

Gemini GMOS-S \citep{Hook04} observations for J0252--0503 \citep[previously described by][]{Yang19} were performed in two wavelength setups both with the R400 grating to cover the small wavelength gaps between detectors, with one setup centered at 860\,nm and the other centered at 870\,nm. These two setups yields a wavelength coverage of 0.6--1.1 $\mu$m and spectral resolution of  $R\sim1300$. Each setup was exposed for an hour. The GMOS data were also reduced with {\tt PypeIt}. The spectra were flux calibrated with the sensitivity function derived from flux standard star GD71 and telluric absorption was corrected using the same method as the NIRES data reduction. 
\edit1{In order to combine the NIRES and GMOS spectra, we scaled the NIRES co-added spectrum to the GMOS flux level using the median in the overlapping wavelength region from 9800 to 10200 \AA. The flux level of the NIRES spectrum is only $\sim$6\% lower than that of GMOS spectrum and the shapes of these two spectra are perfectly matched. Finally, we computed the stacked spectrum in the overlap region after binning the NIRES spectrum to the GMOS wavelength grid.}

Since the flux calibration is crucial for the damping wing analyses, we also obtained near-infrared \emph{Y}, \emph{J}, \emph{H}, and \emph{K}-band photometry with UKIRT/WFCam on 2018 November 27. The on-source times were 8 min in each band. The data were processed using the standard VISTA/WFCAM data-flow system by M. Irwin \citep{Irwin04}. The magnitudes of J0252--0503 were measured to be \emph{Y}=20.33$\pm$0.07, \emph{J}=20.19$\pm$0.07, \emph{H}=20.02$\pm$0.07, and \emph{K}=19.92$\pm$0.08. We then scaled the combined NIRES and GMOS spectrum by carrying out synthetic photometry on the spectrum using the WFCAM \emph{J}-band filter response curve to match the \emph{J}-band photometry for absolute flux calibration. The magnitudes measured from the \emph{J}-band scaled spectrum in the \emph{Y}, \emph{H}, and \emph{K} bands are 20.36, 20.09, and 19.93 mag, respectively. The consistency of magnitudes derived from the fluxed spectrum and UKIRT observations indicates that the spectrophotometric calibration of the spectrum is accurate to within 10\%. Finally, we corrected for Galactic extinction using the dust extinction map derived by \cite{Schlegel98}. The spectrum was then de-redshifted with the systemic redshift $z=7.000\pm0.001$, derived from the IRAM NOrthern Extended Millimeter Array (NOEMA) observations of the far-infrared [\Cii] emission line\footnote{The host galaxy properties of J0252--0503 will be published separately together with [\Cii] observations of a sample of $z>6.5$ quasars.}. The final spectrum used for the following analyses is shown in Figure \ref{fig_spec}. Note that in Figure \ref{fig_spec}, the spectrum is plotted after being rebinned to 200 $\rm km~ s^{-1}$ pixels.

\section{Rest-Frame UV Properties and Black Hole Mass}\label{sec_bh}
In order to derive the rest-frame ultraviolet (UV) properties of J0252--0503, we fit a pseudo-continuum model which includes a power-law continuum, iron (\Feii\ and \Feiii) emission \citep{Vestergaard01,Tsuzuki06}, and Balmer continuum \citep[e.g.][]{Derosa14} to the line-free region of the calibrated and deredshifted spectrum. This pseudo-continuum model is then subtracted from the quasar spectrum, leaving a line-only spectrum. We then fit the \Mgii\ broad emission line in the continuum-subtracted spectrum with two Gaussian profiles. To estimate the uncertainties of our spectral measurements, we use a Monte Carlo approach \citep[e.g.][]{Shen19} to create 100 mock spectra by randomly adding Gaussian noise at each pixel with its scale equal to the spectral error at that pixel. We then apply the exact same fitting algorithm to these mock spectra. The uncertainties of measured spectral properties are then estimated based on the  16\% and 84\% percentile deviation from the median. 
 
The pseudo-continuum model is shown in Figure \ref{fig_spec} and an enlargement of the \Mgii\ region fitting is shown in the right insert panel of Figure \ref{fig_spec}. The fitting procedure yields a power-law continuum of $f_\lambda \propto \lambda^{-1.67\pm0.04}$, from which we measure the rest-frame 3000 \AA\ luminosity to be $\lambda L_{\rm3000\text{\normalfont\AA}}$=(2.5$\pm$0.2)$\times10^{46}$ erg s$^{-1}$, implying a bolometric luminosity of  $L_{\rm bol}$=5.15$\times$ $\lambda L_{\rm3000 \text{\normalfont\AA}}$ = (1.3$\pm$0.1)$\times$10$^{47}$ erg s$^{-1}$ \citep{Shen11}. The rest-frame 1450 \AA\ magnitude is measured to be $M_{1450}=-26.63\pm0.07$. The full width at half maximum (FWHM) and equivalent width (EW) of the \Mgii\ line are measured to be FWHM$\rm _{MgII}=3503\pm205$ km s$^{-1}$ and EW$_{\rm MgII}$=$18.83\pm 0.92$ \AA, respectively. The \Mgii\ emission line is blueshifted by $\Delta_{v, {\rm MgII}}= (712\pm50)~ {\rm km~s^{-1}}$ relative to the systemic redshift determined from the [\Cii] line, similar to other luminous $z\sim7$ quasars in which \Mgii\ blueshifts range from a few hundred to $\sim1000$ km s$^{-1}$ \citep[e.g.][]{Venemans16, Mazzucchelli17, Banados18, Decarli18}.

We adopt the empirical relation obtained by \cite{Vestergaard09} to estimate the black hole mass of J0252--0503, which yields $\rm M_{BH}= (1.39\pm0.16)\times10^{9}~ M_\odot$. \edit1{Note that the quoted black hole mass uncertainty does not include the systematic uncertainties of the scaling relation, which could be up to $\sim0.5$ dex \citep{Shen13}.} By comparing the bolometric luminosity estimated above with the Eddington luminosity, which is $L_{\rm Edd}=1.3\times10^{38}\times {\rm M_{BH}}$, we measure the Eddington ratio of J0252--0503 to be $\lambda_{\rm Edd}=0.7\pm0.1$. Note that the uncertainty quoted here does not consider the systematic uncertainties introduced by both single epoch BH mass estimators and monochromatic bolometric corrections. The Eddington ratio of J0252--0503 is slightly lower than that of the other three luminous $z\ge7$ quasars: $\lambda_{\rm Edd}=1.5^{+0.5}_{-0.4}$ for J1342+0928 \citep{Banados18} at $z=7.54$, $\lambda_{\rm Edd}=1.2^{+0.6}_{-0.5}$ for J1120+0641 at $z=7.09$ \citep{Mortlock11}, and $\lambda_{\rm Edd}=1.25\pm0.19$ for J0038--1527 at $z=7.02$ \citep{Wang18}. If J0252--0503 has been accreting at such Eddington ratio since $z\sim20$ with a radiative efficiency of 10\%, it would require a seed BH of $\sim10^5~{\rm M_\odot}$, which significantly exceeds the predicted mass range from stellar remnant BHs and requires more exotic seed formation mechanisms like direct collapse BHs. 
\edit1{Even if it was accreting at the Eddington limit, J0252--0503 would still require the seed BH to be more massive than $\sim10^4~{\rm M_\odot}$. This indicates that J0252--0503 is one of the few quasars that put the most stringent constraints on SMBH formation and growth mechanisms.}

\begin{figure*}[tbh]
\centering
\includegraphics[width=1.0\textwidth]{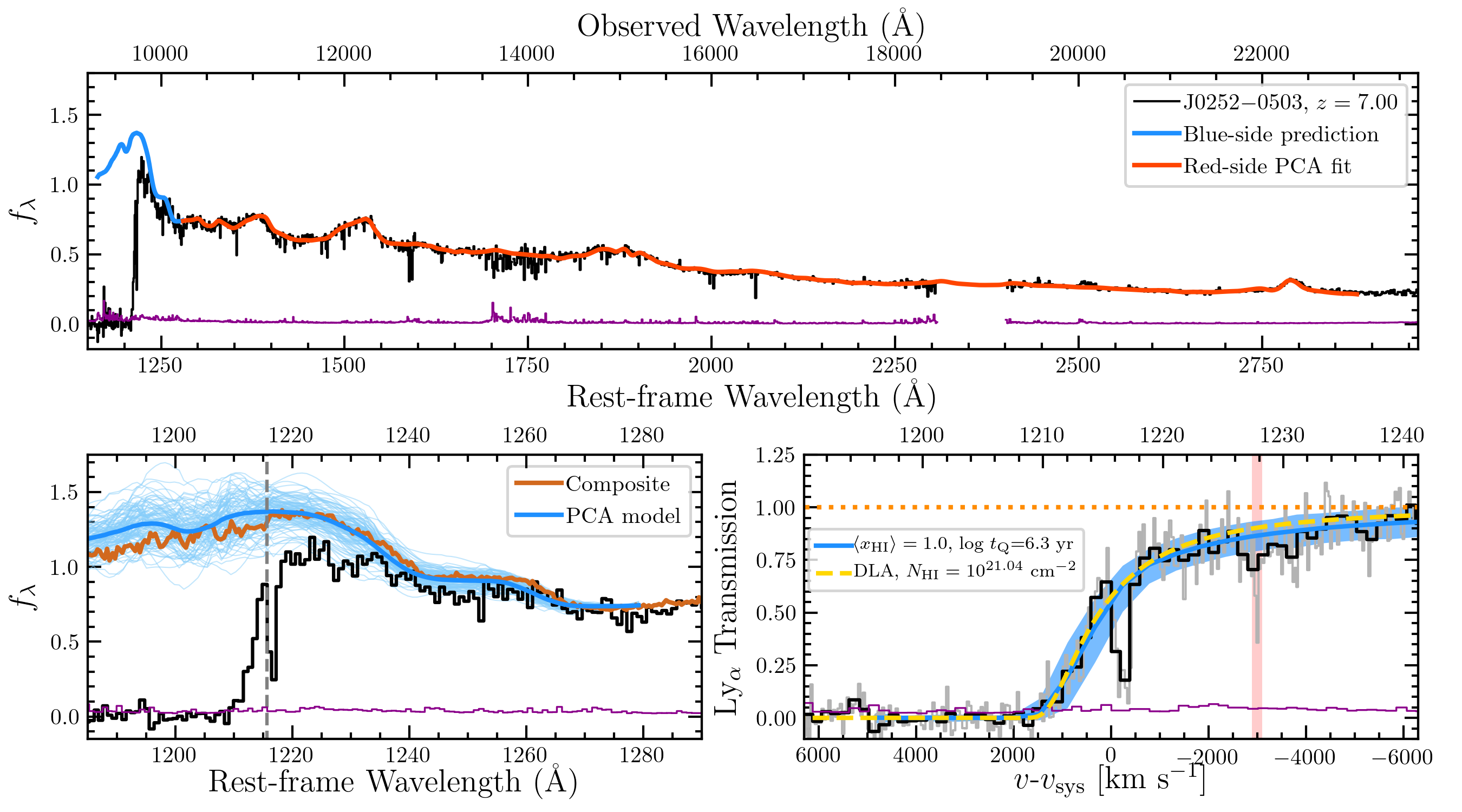}
\caption{
Top: Gemini/GMOS $+$ Keck/NIRES spectrum of J0252--0503, the same as shown in Figure \ref{fig_spec}. The red-side PCA fit and the blue-side prediction are overlaid as red and blue curves, respectively.  
Bottom left: Zoom-in of the Ly$\alpha$ region. The brown and blue lines represent the composite spectrum, and PCA blue-side prediction, respectively. The thinner blue lines show 100 draws from the covariant blue-side prediction error calibrated from the 1\% of quasars that are most similar in the PCA training set. The composite spectra agree well with the PCA prediction, which implies that the detection of a strong damping wing is robust. 
Bottom right: Transmission spectrum of J0252--0503 (the spectrum is normalized by the PCA model). The re-binned spectrum is shown as thick black line, while the un-binned spectrum is shown as a gray line.  The blue solid curve shows the mean transmission spectrum of mock spectra with $\langle x_{\rm HI} \rangle = 1.0$ and $t_{\rm Q}=10^{6.3}$ yr, while the associated blue shaded region shows the 16th--84th percentile range for mock spectra with the above parameters. As a comparison, the transmission spectrum of a DLA model with column density of $N_{\rm HI}= 10^{21.04}~ {\rm cm^{-2}}$ at $z=6.94$, is plotted as a yellow dashed line. The metal line \Alii\ $\lambda1670$ from the $z=4.8793$ absorption system is highlighted by a red transparent vertical line. 
\label{fig_pca}}
\end{figure*}

\section{A Strong Ly$\alpha$ Damping Wing at $z=7$}\label{sec_dp}
Among the six public known $z\ge7$ quasars, two objects already have had damping wing analyses performed \citep{Mortlock11, Bolton11, Bosman15, Greig17, Greig19, Banados18, Davies18b,Durovcikova19}. Two other quasars are too faint ($M_{1450}\gtrsim-25$) for damping wing analyses with current facilities \citep{Matsuoka19a, Matsuoka19b}, and another is a broad absorption line (BAL) quasar in which strong absorption precludes determination of the intrinsic quasar spectrum \citep{Wang18}. Thus, J0252--0503 is the only known bright, non-BAL quasar at $z\ge7$ of which a damping wing analysis has not been performed yet. In order to examine whether the damping wing is present in the spectrum of J0252--0503, we need to know the intrinsic quasar spectrum in the Ly$\alpha$ region (i.e. before IGM attenuation). In the past few years, several methods have been proposed for constructing the quasar intrinsic spectra, including stacking of low-redshift quasar spectra with similar emission line properties \citep[e.g.][]{Mortlock11, Simcoe12, Banados18}, using the principal component analysis (PCA) decomposition approach \citep{Davies18b, Davies18c}, constructing the covariant relationships between parameters of Gaussian fits to Ly$\alpha$ line and those of Gaussian fits to other broad emission lines \citep{Greig17,Greig19}, and using the neural network method \citep{Durovcikova19}. In this paper, we adopt both the empirical composite method and the PCA method to construct the intrinsic spectrum for J0252--0503 as detailed below. 

\subsection{Empirical Composite Spectra from Analogs}\label{subsec_composite}
Since there is a lack of spectral evolution of quasars from low redshifts to high redshifts \citep[e.g.][]{Shen19}, the large sample of SDSS/BOSS quasars at lower redshifts provides a good training set for constructing a high-redshift quasar intrinsic spectrum. First, we use a composite spectrum constructed from a sample of low-redshift quasar analogs to model the intrinsic spectrum. Because the \Civ\ line properties, and especially the line's blueshift, appear to be strongly connected with differences in the quasar spectral energy distribution \citep[e.g.][]{Richards11}, we select quasar analogs from SDSS/BOSS DR14 quasar catalog \citep{Paris18} by matching the \Civ\ blueshifts to J0252--0503.  As most SDSS/BOSS quasars do not have [\Cii] redshifts, we measure the relative blueshifts between the \Civ\ and \Mgii\ lines. This limits us to selecting quasars in the redshift range $2.0<z<2.5$ in order to get Ly$\alpha$, \Civ, and \Mgii\ line properties from BOSS spectra. We also excluded quasars marked as BAL and those without \Mgii\ redshift measurements in the catalog. This yields 85,535 quasars in total. 

Before measuring the line properties from these quasars, we first fit a power-law continuum to the quasar spectrum and subtract it from the data. Instead of fitting the \Civ\ and \Mgii\ lines directly, we use a more robust non-parametric scheme proposed by \cite{Coatman16} to measure the line centroids of \Civ\ and \Mgii\ lines from the continuum subtracted spectra. The relative blueshift between these two lines is then defined as 
\begin{equation}
\Delta v = c\times \left( \frac{1549.06-\lambda_{\rm half, CIV}}{1549.06} - \frac{2798.75-\lambda_{\rm half, MgII}}{2798.75} \right),
\end{equation}
where $c$ is the speed of light and $\lambda_{\rm half, CIV}$ ($\lambda_{\rm half, MgII}$) is the rest-frame wavelength that bisects the cumulative total line flux of \Civ\ (\Mgii). 
We applied this procedure to the spectra of both J0252--0503 and the 85,535 SDSS/BOSS quasars. The blueshift in J0252--0503 is measured to be 4090 km $\rm s^{-1}$. We then select quasars with blueshifts between 3,000 km $\rm s^{-1}$ and 5,000 km $\rm s^{-1}$ and mean spectral signal-to-noise ratios (SNRs) per pixel in the \Civ\ and \Mgii\ regions greater than 4 and 2, respectively. These SNR limits were chosen to yield enough sight-lines to compute a composite. After this, we visually inspected the continuum normalized spectra and removed quasars that have BAL features, proximate damped Ly$\alpha$ systems (PDLAs) and strong intervening absorbers on top of the emission lines. We also reject objects with \Mgii\ line measurements that are strongly affected by sky line residuals and remove targets that have strongly different \Civ\ and \Mgii\ line profiles than J0252--0503 (objects were removed if the line peaks differ by more than three times the spectrum error vector of J0252--0503). In the end, our master quasar analog sample consists of 83 SDSS/BOSS quasars.

Before constructing the composite spectrum, each spectrum was divided by its best fit power-law continuum. Each spectrum was weighted by the average SNR of that spectrum when computing the composite. Then we multiplied the power-law fit from J0252--0503 with the constructed continuum normalized composite, obtaining the composite spectrum shown in Figure \ref{fig_spec}. In order to understand the uncertainties of the composite spectrum and minimize the bias introduced by visual checks, we resampled our parent sample with bootstrapping to construct another 100 composites which are shown as thin orange lines in the insert panel of Figure \ref{fig_spec}. Overall, the constructed composite matches the J0252--0503 spectrum very well across the whole spectral range, except for the Ly$\alpha$ line region. From the left inset of Figure \ref{fig_spec}, we can clearly see that these composites have higher fluxes redward of the Ly$\alpha$ emission line (from 1216\AA\ to 1250\AA\ in rest-frame) than J0252--0503, indicating strong absorption  in the spectrum of J0252--0503.

\subsection{Principal Component Analysis}\label{subsec_pca}
Strong correlations between various broad emission lines of quasars from the rest-frame ultraviolet to the optical are known to exist \citep[e.g.][]{Richards11}. Taking this into account, in principle one can predict the shape of the Ly$\alpha$ line based on the properties of other broad emission lines. \cite{Davies18c} developed a PCA predictive approach based on a training set of $\sim13,000$ quasar spectra from SDSS/BOSS quasar catalog \citep{Paris17} to predict the ``blue-side'' (rest-frame 1175--1280\,\AA) quasar spectrum from the ``red-side'' (rest-frame 1280--2850\,\AA) spectrum. In brief, we performed a PCA decomposition of the training set truncated at 10 red-side and 6 blue-side basis spectra for each quasar. Then we derived a projection matrix relating the best-fit coefficients in the red-side and a template redshift to the coefficients in the blue-side \citep{Suzuki05,Paris11}. With this matrix, we can then predict the blue-side coefficients and thus the blue-side spectrum from a fit to the red-side coefficients and template redshift of a given quasar spectrum. 

We quantify the uncertainties of this prediction by testing the full predictive procedure on every quasar in the training set and computing their relative continuum error \cite[See][for more details]{Davies18c}. We assume a multivariate Gaussian distribution for the relative continuum error, with the covariance matrix determined from the prediction errors measured for similar quasars, i.e., the 1\% nearest neighbors, as the uncertainties of the prediction. 

The advantage of this PCA method compared to the composite spectrum discussed in \S \ref{subsec_composite} is that the PCA approach takes into account the properties of all broad emission lines in the red side rather than just the properties of the \Civ\ line. In addition, we can quantify uncertainties in the blue-side spectrum predictions by testing the method on the input training set. 

In the upper panel of Figure \ref{fig_pca}, we show the red-side PCA fit and blue-side prediction for J0252--0503 on top of the GMOS+NIRES quasar spectrum. In the bottom left panel of Figure \ref{fig_pca}, we show a zoom-in of the Ly$\alpha$ region overlaid with both the blue-side PCA model and the composite spectrum constructed in \S \ref{subsec_composite}. From this zoomed-in plot, we can see that the intrinsic quasar spectrum predicted by the PCA model agrees very well with the composite spectrum. Both models suggest that there is a strong damping wing absorption imprinted on the Ly$\alpha$ emission line of the quasar. Since these two models are consistent with each other, we will only use the PCA continuum model for the following analyses so that we can make use of its well quantified uncertainties.  

\begin{figure}
\centering
\includegraphics[width=0.47\textwidth]{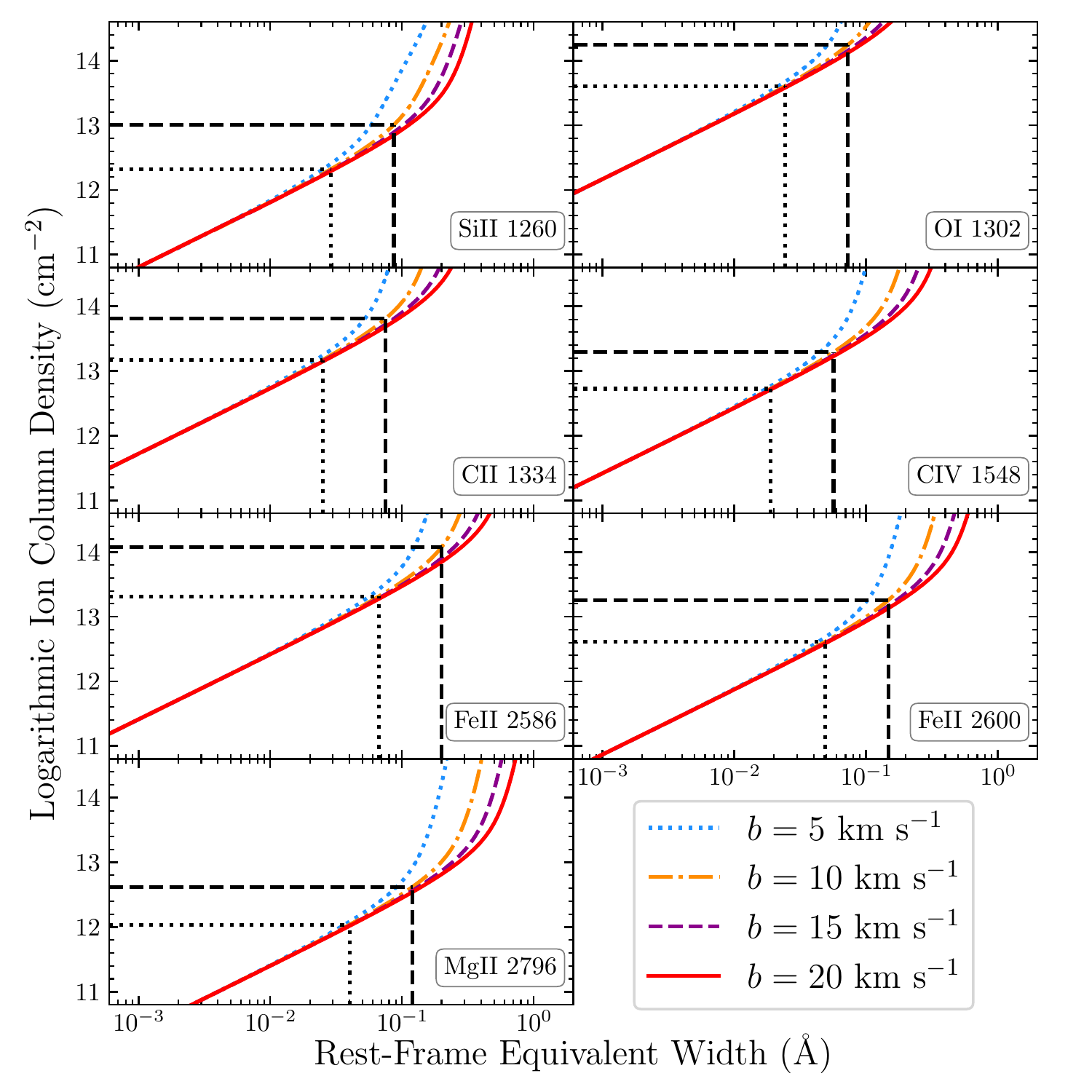}
\caption{Curve of growth analysis to derive column densities for the selected ions. For each panel, the 1$\sigma$ and 3$\sigma$ limits to the equivalent width and column density are shown with dotted and dashed lines, respectively. The colored curves represent $b$-parameters of 5, 10, 15 and 20 $\rm km~s^{-1}$. 
\label{fig_cog}}
\end{figure}

\subsection{Modeling the Damping Wing as a Single DLA}\label{subsec_dla}
The smooth damped absorption profile can be imprinted by either an intervening high column density gravitationally bounded DLA system ($N_{\rm HI}>10^{20}~{\rm cm}^{-2}$) or substantially neutral gas in the IGM. However, DLA systems in the quasar vicinity are very rare at high redshifts. Among more than 250 known $z\gtrsim5.7$ quasars, only a few of them have been identified to be associated with such absorbers close to the quasar redshifts \citep[e.g.][]{DOdorico18, Banados19, Davies19, Farina19}, suggesting that the probability of the strong redward absorption seen in J0252--0503 being caused by a DLA is low. DLA systems are usually associated with a number of metal lines such as \Siii\ $\lambda1260, \lambda1304,\lambda1526$, \Oi\ $\lambda1302$, \Cii\ $\lambda1334$, \Civ\ $\lambda1548,\lambda1550$, \Mgii\ $\lambda2796,\lambda2803$, and a series of \Feii\ lines. Thus, one way to distinguish a DLA damping wing from an IGM damping wing is to search for associated metal absorption features. 

First, we need to determine the redshift of a potential DLA system. To do so, we fit a Voigt profile to the transmission spectrum which is normalized by the PCA continuum model. Since the Doppler parameter, $b$, does not strongly affect the Ly$\alpha$ profile \citep[e.g.][]{Crighton15}, we fixed the $b$ value to be $b=10 {\rm ~km~s^{-1}}$ and use the MCMC sampler \citep[{\tt emcee};][]{emcee} to jointly fit the redshift and \Hi\ column density of a DLA model. During the fit, we masked the narrow absorption at $v\sim0 {\rm ~km~s^{-1}}$. This absorption could be caused by neutral gas inflow since we did not find any associated metal absorption from the quasar spectrum, and it is located at a slightly higher redshift than the quasar if it is caused by neutral hydrogen. The best fit parameters for the system are determined to be $N_{\rm HI}=10^{21.04\pm0.04}~{\rm cm}^{-2}$ and $z_{\rm DLA}=6.939\pm0.002$. In order to qualify the uncertainties of these parameters caused by the continuum model, we then fit DLA models to 100 transmission spectra normalized by the 100 PCA draws shown in the bottom left panel of Figure \ref{fig_pca}. The median values and the mean deviation of 16\% and 84\% percentiles from the median for $N_{\rm HI}$ and $z_{\rm DLA}$ are measured to be $N_{\rm HI}=10^{21.04\pm0.10}~{\rm cm}^{-2}$ and $z_{\rm DLA}=6.940\pm0.003$. To take both the fitting uncertainty and the PCA continuum uncertainty into account, we take $N_{\rm HI}=10^{21.04\pm0.14}~{\rm cm}^{-2}$ and $z_{\rm DLA}=6.940\pm0.004$ as our fiducial parameters for the DLA model, where the uncertainties are the sum of the uncertainties from the {\tt emcee} fitting on the transmission spectrum and the distribution of the 100 draws. This potential DLA system (if it exists) is $\sim2200{\rm ~km~s^{-1}}$ away from the quasar systemic redshift which seems unlikely to be associated with the quasar host galaxy. This best fit DLA model is shown as the yellow dashed line in the bottom right panel of Figure \ref{fig_pca}.  We caution that the resolution of our spectrum is low in the Ly$\alpha$ region, so the DLA fitting procedure might overestimate the $N_{\rm HI}$ if there are some narrow Ly$\alpha$ transmission spikes in the quasar proximity zone that are unresolved in our spectrum. 

We then searched for metal absorption lines at $z\sim6.94$ in the J0252--0503 spectrum. In the end, we did not find any evidence for metal-line absorption at redshifts close to the potential DLA system within $\Delta z \sim \pm0.04$, or $\sim \pm 1500$ km~s$^{-1}$, ten times wider than the redshift uncertainty of the potential DLA system. We also did not find any metal absorption features at the quasar systemic redshift (i.e. from the quasar host galaxy). We then calculated rest-frame equivalent width (EW) 1$\sigma$ limits for each expected metal absorption line as follows:  $W_{\rm r, SiII~1260}\le0.029~{\rm \AA}$, $W_{\rm r, OI~1302}\le0.024~{\rm \AA}$, $W_{\rm r, CII~1334}\le0.025~{\rm \AA}$, $W_{\rm r, CIV~1548}\le0.019~{\rm \AA}$, $W_{\rm r, FeII~2586}\le0.067~{\rm \AA}$, $W_{\rm r, FeII~2600}\le0.049~{\rm \AA}$, $W_{\rm r, MgII~2796}\le0.040~{\rm \AA}$. The EW limits were measured by summing over the normalized pixels over an aperture spanning $\pm2\sigma_{\rm inst}$ from the center of each line, where $\sigma_{\rm inst}=47{\rm ~km~s^{-1}}$ was derived from the NIRES instrumental resolution. In order to derive the column densities for the selected iron, we carried out a curve of growth analysis for four different $b$-parameters following \cite{Simcoe12}. The curve of growth analysis is shown in Figure \ref{fig_cog}. Based on the solar abundance \citep{Lodders03} and the column densities derived by fixing $b=10 {\rm ~km~s^{-1}}$, we find that the metallicity of the potential DLA system is most tightly constrained by \Civ. However, whether high redshift DLAs exhibit \Civ\ is still debated \citep[e.g.][]{DOdorico18,Cooper19}. Thus we use \Mgii\ which sets the second most stringent constraint on the DLA abundance with $\rm [Mg/H]<-4.0$ (3$\sigma$). The DLA abundance $3\sigma$ limits are estimated to be $\rm [Si/H]<-3.6$,  $\rm [O/H]<-3.6$,  $\rm [C/H]<-3.7$,  and $\rm [Fe/H]<-3.3$ based on \Siii\ $\lambda1260$, \Oi\ $\lambda1302$, \Cii\ $\lambda1334$, and \Feii\ $\lambda 2600$, respectively. 
Since the $b$-parameter could be as low as $b=8 {\rm ~km~s^{-1}}$ at high redshifts \cite{DOdorico18}, we also estimate the [Mg/H] based on $b=5 {\rm ~km~s^{-1}}$ and find that $\rm [Mg/H]<-3.7$ (3$\sigma$).

\begin{figure}
\centering
\includegraphics[width=0.47\textwidth]{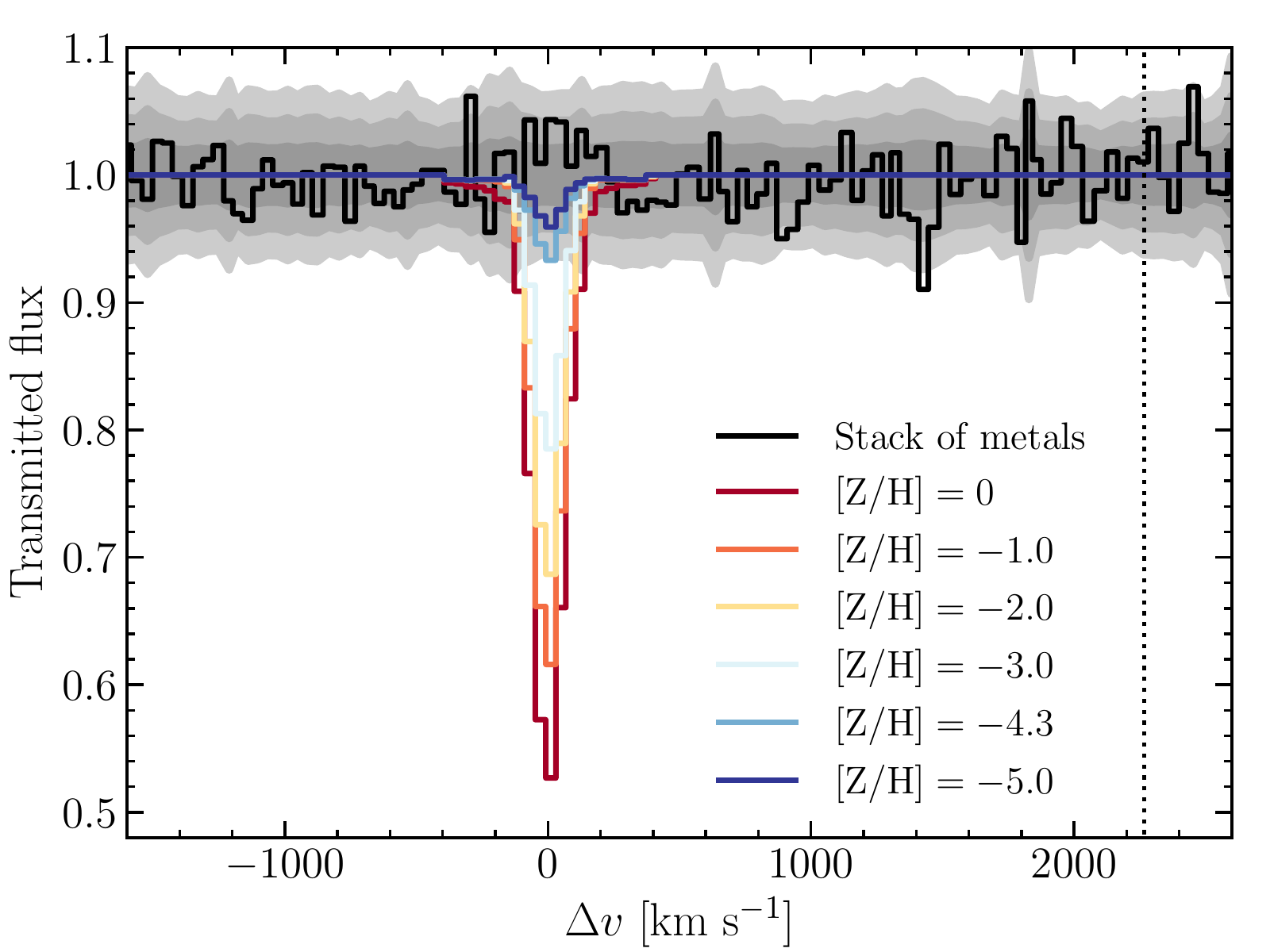}
\caption{Composite stack of heavy-element transitions (\Oi\ 1302, \Cii\ 1334, \Siii\ 1260, \Civ\ 1548, \Feii\ 2586, \Feii\ 2600, and \Mgii\ 2796) generated using an inverse-variance weighted mean for a DLA system at $z=6.94$. The shaded grey regions denote the 1-, 2-, and 3-$\sigma$ error vectors. The quasar systemic redshift is indicated by a black dotted line. Overlaid curves show predicted metal absorption profiles for a DLA with $N_{\rm HI}=10^{21.04}$, $b=10{\rm ~km~s^{-1}}$ and a range of metallicities. The stack shows no statistically significant absorption, suggesting that the metallicity of the absorption system would be more than 10,000 times lower than solar if the damped absorption was produced by a single-component DLA system. 
\label{fig_metal}}
\end{figure}

To further investigate the properties of a possible DLA, we compute the composite stack of the heavy-element transitions shown in Figure \ref{fig_cog} by assuming that there is a metal-poor DLA at $z_{\rm DLA}=6.94$. We stacked the transmitted flux at the expected wavelength using an inverse-variance weighted mean. The composite stack of metal lines is shown in Figure \ref{fig_metal} which shows no significant absorption within $\Delta v \sim 1500 {\rm ~km~s^{-1}}$. Note that the absorption feature at  $v\sim1450~{\rm km~s^{-1}}$ in the stack is caused by the \Feii\ 2344 transition from a $z=3.5425$ absorber (see below). The 1$\sigma$ limit for the dimensionless equivalent width, $W=W_{\lambda} / \lambda$ \citep{Draine11}, for the stack is measured to be $W\le7.3\times10^{-6}$ ($1\sigma$). This corresponds to a limit of $\rm [O/H]<-4.1$ ($3\sigma$) after scaling it to the cross-section and relative abundance of \Oi.
We also compute a set of DLA models by adapting $b=10{\rm ~km~s^{-1}}$ and solar abundance pattern \citep{Lodders03} with varying metallicities. The DLA transmission spectra are computed in the same wavelength grid and same resolution as the spectrum of J0252--0503. The composite stack of these DLA models for different metallicities is also over-plotted in Figure \ref{fig_metal}. The composite stack with $\rm [Z/H]<-4.3$ matches the observed stack at $3\sigma$ level, consistent with our curve of growth analysis of the observed composite metal transitions within 0.2 dex. 
From Figure \ref{fig_cog}, we note that most of the metal transitions are in the linear region of the curve of growth unless $b\lesssim 5 {\rm~km~s^{-1}}$. Thus the metallicity constraint does not change too much by varying the $b$-parameter. By varying $b$ from $5 {\rm~km~s^{-1}}$ to $20 {\rm~km~s^{-1}}$, we can constrain the metallicity of the potential DLA system to be $\rm [Z/H]<-4.5\sim-4.0$. Our analysis indicates that this potential DLA system would be among the most metal-poor DLA systems known \citep[e.g.][]{Cooke11, Banados19}. This suggests that the strong damped absorption is very unlikely to be caused by a DLA.

In addition, we also searched for absorbers at lower redshifts to make sure that the damped Ly$\alpha$ absorption is not contaminated by lower redshift absorbers. We identify five strong \Mgii\ absorption systems at $z=4.8793$, $z=4.7144$, $z=4.2095$, $z=4.0338$, and $z=3.5425$. These systems also exhibit associated \Feii\ lines. The $z=4.8793$ system also has associated  \Alii\ $\lambda1670$ absorption line which falls into the damped absorption region which is masked in the following damping wing analysis. However, this line is very narrow (see the bottom right panel of Figure \ref{fig_pca}) and thus would not be responsible for the smoothed absorption profile on much larger scales. These analyses indicate that the damping wing absorption in the J0252--0503 spectrum is more likely to be imprinted by the neutral IGM rather than by a DLA system or other intervening absorbers, especially considering the fact that J1120+0641 and J1342+0928 also have similar (though slightly weaker) absorption profiles that are not associated with metals \citep[e.g.][]{Simcoe12, Banados18}.

\begin{figure}
\centering
\includegraphics[width=0.47\textwidth]{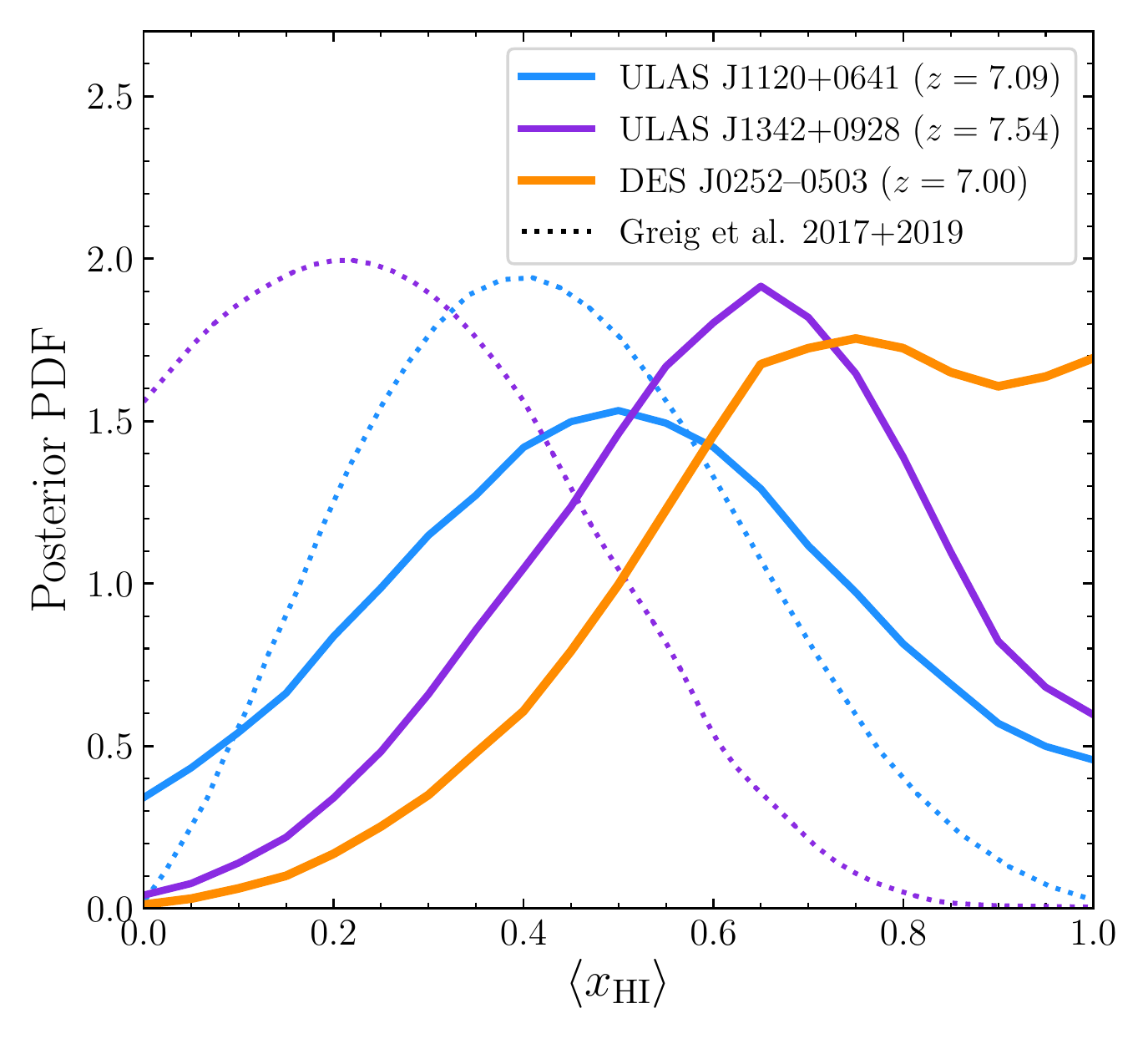}
\caption{Posterior PDFs of  $\langle x_{\rm HI} \rangle$ for all three $z\ge7$ quasars with reported damping wings. The solid orange line denotes J0252--0503. The solid magenta and solid blue lines denote J1342+0928 and J1120+0641 \citep{Davies18b}, respectively. The dotted magenta and dotted blue lines represent the analyses for J1342+0928 and J1120+0641 from \cite{Greig19}, and \cite{Greig17}, respectively. The PDFs for J0252--0503 and the analyses from \cite{Davies18c} are marginalized over quasar lifetime assuming a flat prior covering our entire model grid ($10^3 {\rm yr} < t_{\rm Q} < 10^8 {\rm yr}$).
\label{fig_pdf}}
\end{figure}

\subsection{Constraints on the IGM Neutral Fraction from A Strong Damping Wing at $z=7$}\label{subsec_xhi}
In order to quantitatively assess the damping wing strength and constrain the volume-averaged neutral hydrogen fraction at $z=7$, we applied the methodology from \cite{Davies18b} to this quasar sight-line. We refer the reader to \cite{Davies18b} for a detailed description. In brief, we model the reionization-era quasar transmission spectrum with a multi-scale hybrid model. This model combines large-scale semi-numerical reionization simulations around massive dark matter halos computed in a (400 Mpc)$^3$ volume with a modified version of {\tt 21cmFAST} \citep[][Davies \& Furlanetto in prep]{Mesinger11}, density, velocity, and temperature fields of 1200 hydrodynamical simulation skewers from a separate (100 Mpc/$h$)$^3$ {\tt Nyx} hydrodynamical simulation \citep{Almgren13, Lukic15}, and 1D ionizing radiative transfer which models the ionization and heating of the IGM by the quasar \citep{Davies16}. We then construct realistic forward modeled representations of quasar transmission spectra after accounting for the covariant intrinsic quasar continuum uncertainty from the PCA training. Finally, we use a Bayesian statistical method to recover the joint posterior probability distribution functions (PDFs) of $\langle x_{\rm HI} \rangle$ based on these mock transmission spectra.

The damping wing strength not only depends on the $\langle x_{\rm HI} \rangle$, but also strongly depends on the quasar lifetime, $t_{\rm Q}$, due to the ionization of pre-existing neutral hydrogen along the line of sight by the quasar. In order to measure $\langle x_{\rm HI} \rangle$, we conservatively explore a very broad $t_{\rm Q}$ range with a flat log-uniform $t_{\rm Q}$ prior covering $10^3 {\rm yr} < t_{\rm Q} < 10^8 {\rm yr}$. We then compute the posterior PDF for $\langle x_{\rm HI} \rangle$ by marginalizing over the entire model grid of $t_{\rm Q}$, which is shown in Figure \ref{fig_pdf}. The peak of the PDF leans to the high $\langle x_{\rm HI} \rangle$ end. This is consistent with what we have seen in Figure \ref{fig_pca}, where we show a quasar transmission spectrum model within a $\langle x_{\rm HI} \rangle=1.0$ IGM with a quasar lifetime of $ t_{\rm Q} = 10^{6.3}$ yr. The median and the central 68\% (95\%) confidence interval for $\langle x_{\rm HI} \rangle$ are estimated to be $\langle x_{\rm HI} \rangle = 0.70^{+0.20}_{-0.23} (^{+0.28}_{-0.48})$ from the posterior PDF. As a comparison, we also show the PDFs from the other two $z>7$ quasar sight-lines in Figure \ref{fig_pdf}. Although the redshift of J0252--0503 is lower than the other two quasars, the damping wing in J0252--0503 is the strongest one.

\begin{figure}
\centering
\includegraphics[width=0.47\textwidth]{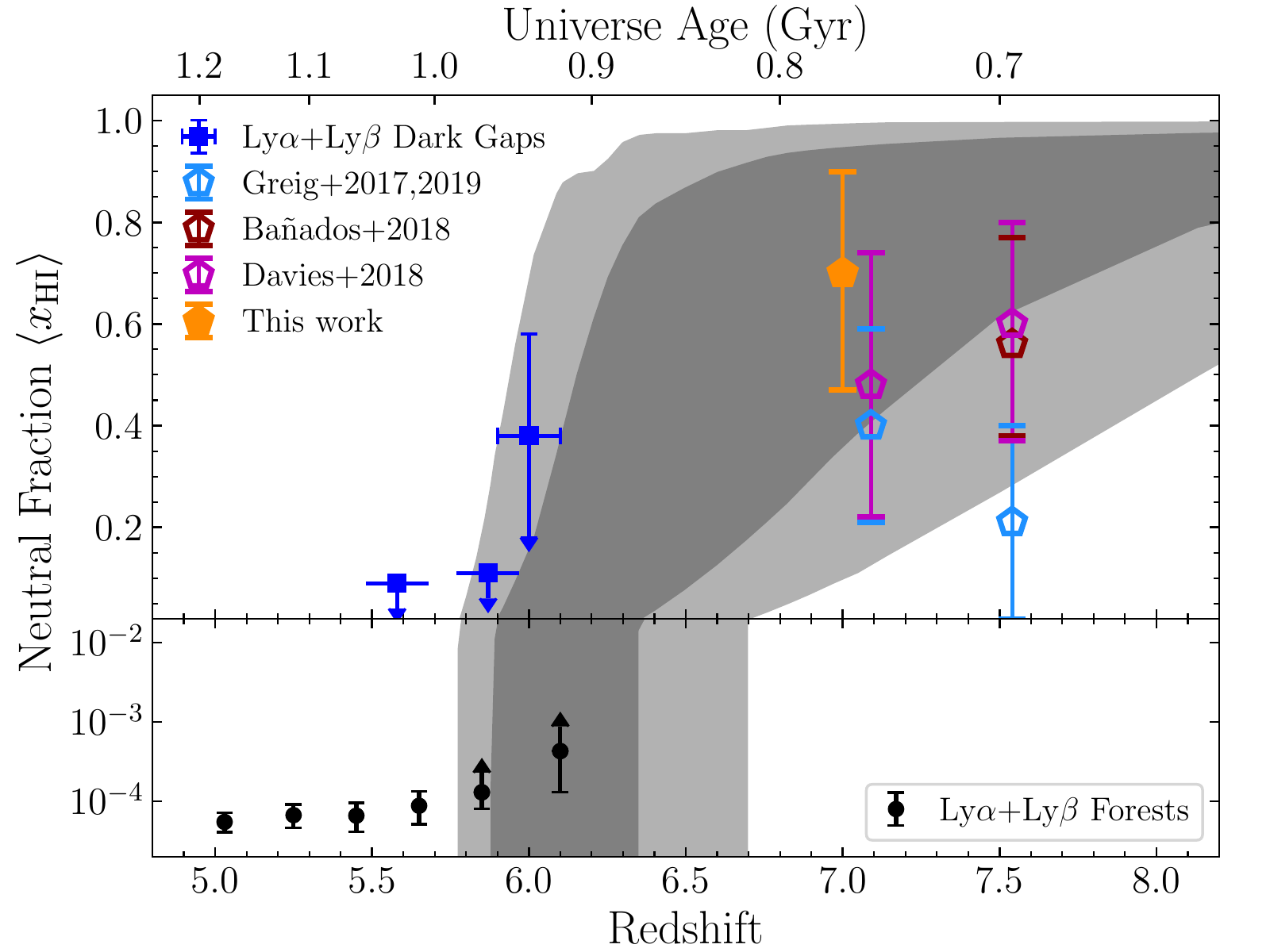}
\caption{Cosmic reionization history constraints from quasar spectroscopy and Planck observations \citep{Planck18}, with the dark and light grey shaded regions corresponding to the 68\% and 95\% credible intervals, respectively. 
Constraints from quasar damping wings are shown as pentagons, with the orange solid pentagon denotes our new measurement with quasar J0252--0503. Also shown are constraints from the Ly$\alpha$+Ly$\beta$ forest dark gaps \citep[blue squares; ][]{McGreer15}, and the Ly$\alpha$+Ly$\beta$ forest opacity \citep[black circles;][]{Fan06}.
\label{fig_xhi}}
\end{figure}

In Figure \ref{fig_xhi}, we plot the $\langle x_{\rm HI} \rangle$ constraints from all three quasar damping wings. In this figure, we also show the $\langle x_{\rm HI} \rangle$ constraints from the Ly$\alpha$+Ly$\beta$ forest \citep{Fan06}, as well as Ly$\alpha$+Ly$\beta$ dark gaps \citep{McGreer15}. All three $z\ge7$ quasars for which a damping wing analysis can be done with current facilities and methodology show evidence of damping wing absorptions, suggesting that the IGM is substantially neutral at $z\ge7$. These constraints are consistent with the integral constraints of $\langle x_{\rm HI} \rangle$ measured from the electron scattering optical depth of the CMB \citep{Planck18} shown as the underlying shaded region in Figure \ref{fig_xhi}. They are also in broad agreement with recent calculations \citep[e.g.][]{Robertson15} and simulations \citep[e.g.][]{Kulkarni19} of the cosmic reionization history, as well as constrains from gamma-ray burst (GRB) damping wings \citep{Totani06, Totani16, Greiner09}, the detections of Ly$\alpha$ emissions from high redshift galaxies \citep[e.g.][]{Ouchi10,Mason18}, and Ly$\alpha$ luminosity functions \citep[e.g.][]{Kashikawa06,Konno18}.

\section{Summary and Discussion}\label{sec_sum}
In this paper we present high-quality near-infrared spectroscopic observations of a bright $z=7$ quasar, J0252--0503, to constrain the cosmic reionization with quasar damping wing modeling and the SMBH growth with BH mass and Eddington ratio measurements. 

We measure the mass of the central SMBH to be $\rm M_{BH}= (1.39\pm0.16)\times10^{9}~ M_\odot$ based on the single-epoch virial method. The Eddington ratio of J0252--0503 is measured to be $\lambda_{\rm Edd}=0.7\pm0.1$, slightly lower than that of the other three $z\ge7$ quasars with similar luminosities. If J0252--0503 has been accreting at such Eddington ratio since $z\sim20$ with a radiative efficiency of 10\%, it would require a seed BH of $\sim10^5~{\rm M_\odot}$, which significantly exceeds the predicted mass range from stellar remnant BHs and requires more exotic seed formation mechanisms like direct collapse BHs. J0252--0503, along with the other three luminous $z>7$ quasars hosting billion solar-mass SMBHs, places the strongest constraints on early BH assembly mechanisms.

In order to investigate whether a damping wing is present in the spectrum of J0252--0503, we explored two different methods to construct the intrinsic spectrum of J0252--0503. The Ly$\alpha$ region of a composite spectrum computed from a sample of \Civ\ blueshift-matched low redshift quasar analogs is consistent with the prediction made by a PCA non-parametric predictive approach. Both methods suggest that a strong damping wing absorption is present in the J0252--0503 spectrum. We modeled the damping wing profile produced by either a single component DLA system or a significantly neutral IGM. However, there is no significant detection of metals at the potential DLA system redshift over a wide range of $\pm1500~{\rm km~s^{-1}}$, suggesting that the strong damping wing in the J0252--0503 spectrum is most likely imprinted by a significantly neutral IGM unless the metallicity of the putative DLA is more than 10,000 times lower than the solar metallicity.

To constrain the IGM neutral hydrogen fraction, $\langle x_{\rm HI} \rangle$, at $z=7$ with the damping wing in J0252--0503, we applied the hybrid model developed by \cite{Davies18b} to our PCA continuum prediction for J0252--0503. Our analysis shows that the damping wing in J0252--0503 is the strongest one yet seen in $z\ge7$ quasar spectra. By marginalizing over quasar lifetime with a log-uniform prior in the range of $10^3 < t_{\rm Q} < 10^8$ yr, we measure the median and the central 68\% (95\%) confidence interval for $\langle x_{\rm HI} \rangle$ to be $\langle x_{\rm HI} \rangle = 0.70^{+0.20}_{-0.23} (^{+0.28}_{-0.48})$ at $z\sim7$. 
\edit1{The recent study by \cite{DAloisio20} suggests that unrelaxed gaseous structures may exist in the post-reionization IGM, meaning that the mean free path of ionizing photons is shorter compared with a model that assumes the gas is fully relaxed. The mean free path in the quasar proximity zone, however, should still be quite long due to the strong ionizing radiation of the central luminous quasar \citep{McQuinn11,Davies19}. Thus our constraints on $\langle x_{\rm HI} \rangle$ based on damping wing analysis should not be strongly affected by unrelaxed baryons in the proximity zone.}

Despite the limited precision of quasar continuum reconstructions and the degeneracy of $\langle x_{\rm HI} \rangle$ and quasar lifetime,  the damping wing is still highly effective in constraining the reionization history. Although the currently available sample of quasar sight-lines at $z\gtrsim7$ is very small, more luminous $z\gtrsim7$ quasars are expected to be found in the next few years through ongoing quasar searches \citep[e.g.][]{Banados18, Wang18, Yang19, Matsuoka19a, Reed19}. Moreover, the {\it Euclid} wide survey will be online soon, and will discover more than 100 quasars at $z>7$ \citep{Barnett19}. In addition, the Near-Infrared Spectrograph (NIRSpec) on the James Webb Space Telescope (JWST) will provide much higher quality spectroscopic data for more precise quasar damping wing analyses. Thus, we expect that quasar damping wing analyses will have the capability to place increasingly strong constraints on the cosmic reionization history during the next several years.

\acknowledgments
Support for this work was provided by NASA through the NASA Hubble Fellowship grant \#HST-HF2-51448.001-A awarded by the Space Telescope Science Institute, which is operated by the Association of Universities for Research in Astronomy, Incorporated, under NASA contract NAS5-26555.
J. Yang, X. Fan and M. Yue acknowledge support from the US NSF Grant AST-1515115 and NASA ADAP Grant NNX17AF28G. 
Research by A.J.B.\ is supported by NSF grant AST-1907290. 
X.-B.W. and L.J. acknowledge support from the National Key R\&D Program of China (2016YFA0400703) and the National Science Foundation of China (11533001 \& 11721303). 
ACE acknowledges support by NASA through the NASA Hubble Fellowship grant \#HST-HF2-51434 awarded  by  the  Space  Telescope  Science  Institute,  which  is  operated  by  the   Association  of  Universities for  Research  in  Astronomy,  Inc.,  for  NASA,  under  contract  NAS5-26555. 
B.P.V. and F.W. acknowledge funding through the ERC grant “Cosmic Gas.”

The data presented in this paper were obtained at the W.M. Keck Observatory, which is operated as a scientific partnership among the California Institute of Technology, the University of California and the National Aeronautics and Space Administration. The Observatory was made possible by the generous financial support of the W. M. Keck Foundation. This work was supported by a NASA Keck PI Data Award, administered by the NASA Exoplanet Science Institute. Data presented herein were obtained at the W. M. Keck Observatory from telescope time allocated to the National Aeronautics and Space Administration through the agency's scientific partnership with the California Institute of Technology and the University of California. This research based on observations obtained at the Gemini Observatory (GS-2018B-FT-202), which is operated by the Association of Universities for Research in Astronomy, Inc., under a cooperative agreement with the NSF on behalf of the Gemini partnership: the National Science Foundation (United States), National Research Council (Canada), CONICYT (Chile), Ministerio de Ciencia, Tecnolog\'{i}a e Innovaci\'{o}n Productiva (Argentina), Minist\'{e}rio da Ci\^{e}ncia, Tecnologia e Inova\c{c}\~{a}o (Brazil), and Korea Astronomy and Space Science Institute (Republic of Korea). UKIRT is owned by the University of Hawaii (UH) and operated by the UH Institute for Astronomy; operations are enabled through the cooperation of the East Asian Observatory.

The authors thank Percy Gomez and Greg Doppmann for their expert support and advice during our NIRES observing runs. Some of the data presented herein were obtained using the UC Irvine Remote Observing Facility, made possible by a generous gift from John and Ruth Ann Evans.

The authors wish to recognize and acknowledge the very significant cultural role and reverence that the summit of Mauna Kea has always had within the indigenous Hawaiian community. We are most fortunate to have the opportunity to conduct observations from this mountain.

\vspace{5mm}
\facilities{Gemini(GMOS), Keck(NIRES), UKIRT(WFCam), NOEMA}

\software{astropy \citep{astropy}, emcee \citep{emcee}, matplotlib \citep{matplotlib}, PypeIt \citep{pypeit}}

\end{document}